\documentclass[12pt,titlepage]{article}

\def \bold0{{\mbox{\boldmath $0$}}}
\def \bold1{{\mbox{\boldmath $1$}}}

\def \boldX{{\mbox{\boldmath $X$}}}

\def \boldbeta{{\mbox{\boldmath $\beta$}}}

\def \boldeps{{\mbox{\boldmath $\epsilon$}}}

\def \var{\hbox{var}}

\usepackage{graphicx}
\usepackage{multirow}
\usepackage{epsf}
\usepackage{amsmath}
\usepackage{caption}
\usepackage{subcaption}

\begin{document}

\baselineskip 22pt
\begin{titlepage}

\title{A Comparison of Location-Effect Identification Methods for Unreplicated Fractional Factorials in the Presence of Dispersion Effects}

\author{Thomas M. Loughin \\
Department of Statistics and Actuarial Science\\
Simon Fraser University\\Burnaby, BC, V5A 1S6, Canada\\[.2 in]
Yan Zhang \\ Risk Assurance Department\\PricewaterhouseCoopers\\
26/F Office Tower A\\Beijing Fortune Plaza\\7 Dongsanhuan Zhong Road\\Chaoyang District\\Beijing 100020, PRC }

\date{\today}

\maketitle
\end{titlepage}

\begin{abstract}
Most methods for identifying location effects in unreplicated fractional factorial designs assume homoscedasticity of the responses.  However, dispersion effects in the underlying process may create heteroscedasticity in the responses.  This heteroscedasticity may go undetected when location-effect identification is pursued; indeed, methods for identifying dispersion effects typically require first modeling location effects.  Therefore, it is imperative to understand how location-effect identification methods function in the presence of undetected dispersion effects.  We use simulation studies to examine the robustness of four different location-identification methods---Box and Meyer (1986), Lenth (1989), Berk and Picard (1991), and Loughin and Noble (1997)---under models with one or two dispersion effects of varying sizes.  We find that the first three methods are perform fine with respect to error rates and power, but the Loughin-Noble method loses control of the individual error rate when moderate-to-large dispersion effects are present.\\[.1 in]

\end{abstract}

\baselineskip 22pt

\section{Introduction}\label{Intro}
In many research and quality-improvement settings, experimental studies are used to test the effects of a number of factors that may have an impact on some measured responses (Dean and Lewis 2006).  Most often, investigators seek ``location effects''---factor main effects or interactions that impact the mean of the response.   In many cases, logistics dictate that the experiment be run without replication, disabling the potential to derive an independent estimate of error for testing these location effects (Box, Hunter, and Hunter 2005, Wu and Hamada 2000).  There is a history to the development of methods for identifying location effects in unreplicated experiments, starting with Daniel (1959).   A general overview and comparison of most of these methods---including those by Daniel (1959), Zahn (1975), Box and Meyer (1986a), Benski (1989), Bissell (1989), Lenth (1989), Berk and Picard (1991), Juan and Pe\~{n}a (1992), and Dong (1993)---is given by Hamada and Balakrishnan (1998).

In the past 20--30 years, more recognition has been given to the possibility that the factorial effects may also impact {\em variance} of the responses (see, e.g., Box and Meyer 1986b, Bergman and Hyn{\'e}n 1997, Brenneman and Nair 2001, McGrath and Lin 2001ab).  Such effects are called ``dispersion effects,'' and they raise or lower the variance corresponding to their different levels, thus creating heteroscedasticy among the responses.  Dispersion effects may be present regardless of whether or not the model used for the analysis accounts for them.


It is known that the heteroscedasticity created by a dispersion effect induces correlation on the location-effect estimates from a linear model (Grego et al.~2000, McGrath and Lin 2003; see details in Section \ref{DispEffs}).  However, except for a single example case mentioned in Pan (1999), it is not currently known how heteroscedasticity might affect any of the above-mentioned location-effect identification methods, either directly or through correlations it induces upon the estimates of location effects.  The goal of this paper is to investigate whether common statistical analysis methods for location effects in unreplicated $2^k$ (fractional) factorial designs are disturbed by the undetected presence of dispersion effects.  Specifically, we use a simulation study to generate data from $2^k$ designs where different numbers and intensities of dispersion effects are present.  We analyze these data using four methods for identifying location effects that are not meant to account for the unequal variances that the dispersion effects create: the Lenth method (Lenth 1989), the Berk and Picard method (Berk and Picard 1991), the Box and Meyer method (Box and Meyer 1986a), and a permutation-based procedure, the Loughin and Noble method (Loughin and Noble 1997).  We measure various forms of power and error rates for each method and summarize the results.

This work is important for two reasons.  First, dispersion effects may lurk when they are unexpected and untested.  Thus, location-effect identification methods need to be robust against their existence, and if they are not, then the risk associated with their use needs to be quantified. Second, when dispersion effects {\em are} expected and tested, most methods for identifying them---e.g., Box and Meyer (1986b), Bergman and Hyn\'{e}n 1997, McGrath and Lin (2001b) and Brenneman and Nair (2001)---assume that the location model is known.  The dispersion effects are then identified using the residuals from an estimated version of the model.  In practice, the true location model is rarely known; instead it is identified using one of the methods mentioned above.  However, these methods are not perfect, and it is well known that failure to properly identify location effects can have adverse effects on dispersion-effect identification (Pan 1999, McGrath and Lin 2001a, Pan and Taam 2002).  If the presence of dispersion effects causes location-effect models to be mis-identified, then this could in turn inhibit the detection of these dispersion effects.


The outline of the paper is as follows.  Section \ref{DispEffs} reviews the impact that dispersion effects have on the OLS estimates from model (\ref{eq:usual}).  Section \ref{LocEffs} provides some details on the location-effect testing methods used in our study. The simulation model and other settings are given in Section \ref{SimDescr}, and the results of the simulations are presented in Section \ref{SimResults}.  We draw conclusions and note a particularly important ramification of the results in Section \ref{concl}.

\section{How Dispersion Effects Change Location-Effect Estimates} \label{DispEffs}

We focus on experimental designs of the $2^k$ structure, because these are popular and well-studied choices for screening large numbers of factors (Wu and Hamada 2000; Box, Hunter, and Hunter 2005).  These designs consist of $k$ factors to be studied, each at two levels.  Fractional factorial designs exist that allow study of more than $k$ factors within the same experimental design construct.  The effect estimates have exactly the same mathematical properties regardless of whether the factorial design is whole or fractional, although interpretation of the effect estimates in the latter case is clouded by the fact that each estimate represents more than one true effect.  Statistical analysis techniques treat all such designs as equivalent, so we refer here only to full $2^k$ factorials for simplicity.

It is common to use a linear model to represent the responses from a $2^k$ design.  Let $n=2^k$ denote the number of experimental runs and let $\mathbf{X}_{n\times n} = [\mathbf{x}_{0},\mathbf{x}_{1},...,\mathbf{x}_{n-1}]$ be the design matrix for an unreplicated $2^k$ factorial experiment, where $\mathbf{x}_{0}=(1,...,1)'$ and $\mathbf{x}_{j}=(\pm{1},...,\pm{1})',j=1,...,n-1$, are pairwise orthogonal. The usual linear statistical model is
\begin{equation}\label{eq:usual}
\mathbf{Y}=\mathbf{X}\boldbeta+ \boldeps
\end{equation}
where $\mathbf{Y}=(y_1,y_2,...,y_n)'$ is the vector of responses (possibly transformed to fit model assumptions), and $\boldbeta_{n\times 1}$ is a vector of unknown parameters.  The parameters $\beta_1,\ldots,\beta_{n-1}$ are called ``location effects.'' Finally, $\boldeps=(\epsilon_1,...,\epsilon_n)'$ is the vector of random error terms. The typical assumption is that $\epsilon_i, i=1,...,n$, are independent and identically distributed (i.i.d.) $N(0,\sigma^2)$ random variables, where $\sigma^2$ is the error variance of the responses.

Due to the to structure of $\boldX$, ordinary least squares (OLS) estimates of $\boldbeta$ are easy to compute and enjoy some useful properties.  In particular, the location-effect estimates, $\hat{\beta}_j,~j=1,\ldots,n-1$ are i.i.d $N(0,\sigma^2/n)$.  However, the lack of replication prevents the calculation of an unbiased estimate of $\sigma^2$ without added assumptions.  As a result, numerous methods, each based on different sets of assumptions, have been developed for testing the significance of the factorial effects, as detailed in Section \ref{Intro}.

If dispersion effects are present in a model, correlations are induced among the OLS estimates $\hat{\beta_j},~j=1,\ldots,n-1$.  In this section we review the nature of these correlations.

To fix ideas, we consider a standard linear model that accounts for heteroscedasticity:
\begin{equation}\label{heteroscedastic}
\mathbf{Y}=\mathbf{X}\boldbeta+\boldeps,\quad\quad \epsilon\sim N(0,\Sigma)
\end{equation}
where $\Sigma$ is a diagonal matrix that depends on the factors through columns of $\boldX$.  There are many ways to relate the covariance matrix to $\boldX$. Most of literature has concentrated on two specific parametric forms: the additive model and the multiplicative model. Rao (1970), Bergman and Hyn\'{e}n (1997), and Brenneman and Nair (2001) use the additive model, $\sigma_{i}^2=\gamma_0+\sum_{j=1}^{n-1}x_{ij}\gamma_j,$ where $\gamma_j$ represents the unknown dispersion effect associated with $\mathbf{x}_j$. Alternatively, Cook and Weisberg (1983), Davidian and Carroll (1987), McGrath and Lin (2001ab, 2003), and Henrey and Loughin (2017) consider a multiplicative variance model in their work.  In this paper, we assume a multiplicative variance model similar to that used by McGrath and Lin (2001ab, 2003), in which the $i$th diagonal element of $\Sigma$ is
\begin{equation}\label{dispEff}
\sigma_{i}^2=\sigma^2\prod_{j=1}^{n-1}\Delta_{j}^{x_{ij}/2},
\end{equation}
where $\sigma^2$ and $\Delta_j,j=1,...,n-1$ are unknown parameters. In particular, suppose that $\Delta_j=1$ for all $j\ne d$.  Then $\Delta_d$ represents the ratio of variance for observations at the $+$ level of $x_d$ to the variance at the $-$ level.  More generally, $\Delta_d$ is the ratio of variances at the $+$ and $-$ levels of $x_d$, holding all other factors constant.

Now suppose that data arise from models (\ref{heteroscedastic}) and (\ref{dispEff}), but location effects are estimated by OLS.  Then $\hat{\boldbeta}$ is distributed as multivariate normal with mean $\boldbeta$ and covariance matrix $\sigma^2(X'\Sigma X)/2^{2k}$. The diagonal entries of this covariance matrix of $\hat{\boldbeta}$ are identical, so that each $\hat{\beta_j}$ has the same marginal distribution. However, these effect estimates have non-zero correlation, so they are no longer independent.


In particular, suppose that factor $A$ produces a dispersion effect, $\Delta_A,$ and that there are no other dispersion effects present. Then for all $j=1,\ldots,n-1$
\begin{equation}\label{disp1}
\var(\hat{\beta}_j)=\sigma^2(\sqrt{\Delta_A}+\frac{1}{\sqrt{\Delta_A}})/2^{k+1}.
\end{equation}
Furthermore, let $j_1$ and $j_2$ be indexes for any two columns of $\boldX$ whose elementwise product is the column for $A$.  As in McGrath and Lin (2001b), we refer to $\{A, j_1, j_2\}$ as an ``interaction triple'' and use the common notation $j_1 \circ j_2 =A$ to represent the relationship.  Grego et al.~(2000) and McGrath and Lin (2003) show that the correlation between $\hat{\beta}_{j_1}$ and $\hat{\beta}_{j_2}$ is
\begin{equation}
\rho(\hat{\beta}_{j_1},\hat{\beta}_{j_2})=
\dfrac{(\sqrt{\Delta_A}-\frac{1}{\sqrt{\Delta_A}})}{(\sqrt{\Delta_A}+\frac{1}{\sqrt{\Delta_A}})}.
\end{equation}
If $j_1$ and $j_2$ do not form an interaction triple with $A$, then the correlation is 0.  Note that there are $n/2 -1$ pairs of columns that satisfy $j_1\circ j_2 =A$ where both $j_1>0$ and $j_2>0$.  Thus, each dispersion effect creates several correlated pairs of location-effect estimates.

These correlations can be fairly large under very realistic circumstances.  For a moderately-sized dispersion effect such as $\Delta_A=9$---i.e., a standard deviation ratio of 3---the correlation is 0.8.  For a larger dispersion effect of 25, the correlation is 0.92.

For multiple dispersion effects, the situation is more complicated. If there are two dispersion effects corresponding to factors $A$ and $B$, with magnitudes $\Delta_A$ and $\Delta_B$ respectively, an extra dispersion effect is induced in their interaction column $AB$ (McGrath and Lin 2003):
$$
\Delta_{AB}=\frac{1+\Delta_A\Delta_B}{\Delta_A+\Delta_B}.
$$
Thus, $\{A,B,AB\}$ form a ``dispersion triple.''  In this setting, the OLS location-effect estimates still have common variance,
$$
\var(\hat{\beta}_i)=\sigma^2(\sqrt{\Delta_A}+\dfrac{1}{\sqrt{\Delta_A}})(\sqrt{\Delta_B}+\dfrac{1}{\sqrt{\Delta_B}})/2^{k+2}.
$$
However, the correlation pattern is more complex (Grego et al. 2000):
\begin{equation}\label{eqn:corr}
\rho(\hat{\beta}_{j_1},\hat{\beta}_{j_2})=\left\{\begin{array}{llll}
\dfrac{(\sqrt{\Delta_A}-\dfrac{1}{\sqrt{\Delta_A}})}{(\sqrt{\Delta_A}+\dfrac{1}{\sqrt{\Delta_A}})}, & \textrm{if $j_1\circ j_2=A$}\\
\dfrac{(\sqrt{\Delta_B}-\dfrac{1}{\sqrt{\Delta_B}})}{(\sqrt{\Delta_B}+\dfrac{1}{\sqrt{\Delta_B}})}, & \textrm{if $j_1\circ j_2=B$}\\
\dfrac{(\sqrt{\Delta_A}-\dfrac{1}{\sqrt{\Delta_A}})(\sqrt{\Delta_B}-\dfrac{1}{\sqrt{\Delta_B}})}{(\sqrt{\Delta_A}+\dfrac{1}{\sqrt{\Delta_A}})(\sqrt{\Delta_B}+\dfrac{1}{\sqrt{\Delta_B}})}, & \textrm{if $j_1\circ j_2=AB$}\\
0 & \textrm{otherwise}\
\end{array}\right.
\end{equation}

Thus, when one dispersion effect exists, every location effect estimate except the one corresponding to the dispersion effect is correlated with one other.  When there are two dispersion effects, the location-effect estimates corresponding to the dispersion triple are pairwise correlated, and every other location effect estimate is pairwise correlated with three other location effect estimates. For the example above, $(\hat{\beta}_C, \hat{\beta}_{AC}, \hat{\beta}_{BC}, \hat{\beta}_{ABC})$ form a ``correlation quadruple''.  For convenience we also would call $(\hat{\beta}_0, \hat{\beta}_A, \hat{\beta}_B, \hat{\beta}_{AB})$ a correlation quadruple, even though we are not generally concerned with testing the intercept.

\section{Location-Effect Testing Methods Used in Simulations}\label{LocEffs}

Grego et al.~(2000) studies the impacts that the dispersion-induced correlations have on the expected order statistics in a half-normal plot in $2^3$ and $2^4$ factorials.  They find that the expected order statistic corresponding to the most extreme effect estimate is slightly closer to 0 than expected under independence, causing the respective effect estimate to seem slightly less important than it should be.  For example, when $\Delta_A=9$, the actual expected order statistics are 5\% and 3\% smaller than their counterparts under independence for the $2^3$ and $2^4$ designs, respectively.  It is not known whether these correlations affect more formal analysis procedures that are based on an assumption of independence of the estimated location effects.  Our simulation study attempts to address this for a variety of well-known procedures.

Hamada and Balakrishnan (1998) review and compare the performance of 24 existing or modified versions of methods for testing for location effects under model (\ref{eq:usual}).  They compare the methods' error rates and powers under a $2^4$ design with varying numbers of effects of equal size. They find that, while some methods perform quite poorly at times, none of the remaining methods are uniformly better than others.  We therefore e selectthree methods from their study that have reasonably good power overall, while representing different foundational structures.


The Lenth (1989) method is a particularly easy method to compute and is recommended by Wu and Hamada (2000).  It standardizes the location-effect estimates contrasts by the estimated \emph{pseudo standard error} (PSE) that is calculated in two steps.  First, any effects with estimated magnitudes more than 3.75 times the median magnitude are temporarily removed.  Then the PSE is set to 1.5 times the median magnitude among the remaining effect estimates, and all estimates are standardized using this PSE.  Lenth suggests comparing standardized effects to a $t_{(n-1)/3}$ distribution, although Loughin (1998) and Ye and Hamada (2000) separately determine alternative critical values that achieve control of various error rates.

Berk and Picard (1991) propose an ANOVA-based method using a trimmed mean squared error (TMSE).  It is similar to the Lenth method in that the smaller effect estimates form the error term against which the larger estimates are compared.  However, the Berk and Picard method uses a fixed number (e.g., 60\%) of the effects in the error term rather than choosing the number adaptively.  It also sums the squared coefficients rather than using a median absolute value. This has the potential to make the error term more stable when the number of inert contrasts is at least as large as the number assumed.  Berk and Picard (1991) obtain critical values based on a numerical study. The critical values given in Table 1 of their paper are computed for samples of sizes $N=8,12,16,20,32$.

In contrast to these methods, Box and Meyer (1986a) suggest a Bayesian approach based on the sparsity assumption that only a few of the factorial effects are are likely to be active. It is assumed {\it a priori} that~$\beta_j=0$ for inactive location effects and $\beta_j\sim N(0,\tau^2)$ for active location effects, $j=1,\ldots,n-1$.  An effect $\beta_j$ is active with prior probability $\eta$.  Thus, estimated contrasts corresponding to inactive effects have distribution $N(0,\sigma^2/2^k)$ while estimated contrasts corresponding to active effects have distribution $N(0,\tau^2 + \sigma^2/2^k)$. In other words, they assume that $\hat{\beta}_1,\ldots,\hat{\beta}_{n-1}$ are i.i.d.~from a scale-contaminated normal distribution. For each effect, the marginal posterior probability of being active is computed by numerical integration.  Box and Meyer (1986a) recommend that the effects whose marginal posterior probability exceeds 0.5 be declared active.  They suggest values for the prior parameters based on analysis of ten published data sets of unreplicated fractional factorial designs, and report that ``the conclusions to be drawn from analysis are usually insensitive to moderate changes'' in these parameters (p.~13).

Finally, we include a fourth method that has a completely different basis from the previous three, the permutation-based procedure of Loughin and Noble (1997).  It is a sequential algorithm that has the potential to test up to $n-k$ effects. As with most randomization tests, $y_1,\ldots,y_n$ are exchangeable under the null hypothesis of no active effects.  The test statistic is the largest magnitude among the location-effect estimates.  Testing whether the corresponding effect is active consists of finding the re-randomization distribution test statistic.  Tests for other effects are constructed in the same manner, except that the permutations are applied to residuals from a model containing all effects of larger estimated magnitude than the one being tested.  The tests are carried out in order of decreasing effect-estimate magnitude, leading to a sequence of p-values.  This sequence is not necessarily monotone, so comparisons are made against simulated critical values in a ``step-up'' manner.  Specifically, the effect of smallest magnitude whose p-value is below the critical value is declared active, as are all effects with larger magnitudes.

Thus, we have four location-effect identification methods based on three very different analysis approaches (standardizing contrasts is a popular technique, so we have two variants of this approach in the Lenth and Berk-Picard procedures).  However, all assume that the estimated effects are independent of one another, which is clearly not true in the presence of dispersion effects.

\section{Description of Simulation Study}\label{SimDescr}
We now describe the simulation study conducted to examine the robustness of these four location-effect identification methods in the presence of one or two dispersion effects. We first provide an overview of the study, then give details on the specific settings.

The general process is as follows.  We generate data from model (\ref{heteroscedastic}) and (\ref{dispEff}) with $k=4$, so that homoscedasticity is the only violated assumption for the testing methods.  We fix the true location effects and true dispersion effects at specific sizes for both models.  We then simulate data sets under these models and apply the four testing methods to each data set.  We separately record the proportion of times that each factorial effect is identified as active and use these to compute power and error rate summaries for each method.  All simulations are run in SAS/IML using 1825 data sets per setting.  This number ensures that any error rate that should nominally be 0.05 will lie between 0.04 and 0.06 with 0.95 probability.

Simulations are conducted under the $2^4$ setting for several reasons. It is the smallest size for which both location and dispersion effects can practically be tested. It is also among the most common designs reported in the literature, and it corresponds to the experiment size reported in the extensive simulation study by Hamada and Balakrishnan (1998). 
Because the properties of OLS estimates in the presence of dispersion effects are the same regardless of the experiment size, we expect that our results should be reasonable representations of what can be expected in other common sizes of experiment.

We use the usual letter identifiers ($A$, $B$, $C$, $D$, $AB$,..., $ABCD$) to label factorial effects.  We denote a location effect model by ${\cal L}$ and a dispersion-effect model by ${\cal D}$.  We consider models with either one dispersion effect---${\cal D}=\{A\}$ with effect size $\Delta_A$---or two---${\cal D}=\{A,B\}$ with effect sizes $(\Delta_A,\Delta_B)$.  In the former case, we consider location models with either 0, 1, or 2 active effects, while in the latter we consider just 0 or 1 active effect.  In all cases, we assign the active location effects strategically as indicated in Table \ref{cases} to allow us to examine different cases where correlated location-effect estimates and/or location-dispersion confounding may be present.

\begin{table}[!htbp]
\begin{center}
\caption{Location and dispersion models used in the simulations.}\label{cases}
\bigskip
\renewcommand{\arraystretch}{1.5}
\begin{tabular*}{0.8\textwidth}{@{\extracolsep{\fill}}c|c|c}
\hline
\multicolumn{1}{c|}{$\cal{D}$}&\multicolumn{1}{c|}{$\cal{L}$}& \multicolumn{1}{c}{Explanation}\\\hline\hline
\multirow{2}{*}{$A$}& $A$ & Location effect is dispersion effect \\
& $B$ & Location effect is different from dispersion effect\\\hline
\multirow{3}{*}{$A$}& $B, C$ & Neither of location effects is dispersion effect\\
& $A, B$ & One location effect is dispersion effect\\
& $B, AB$ & The interaction of location effects is dispersion effect\\\hline

\multicolumn{1}{c|}{$A,B$}& $A$ & Location effect is one of the dispersion effects \\
\multicolumn{1}{c|}{$A,B$}& $C$ & Location effect is not associated with dispersion effects \\
\multicolumn{1}{c|}{$A,B$}& $AB$ & Location effect is the interaction of the dispersion effects \\ \hline \hline
\end{tabular*}
\end{center}
\end{table}

Dispersion-effect sizes are set to values $\Delta_A =1,2^2,3^2,5^2,10^2,20^2,50^2$, representing a range from null to extremely large values.\footnote{It is somewhat easier to measure heteroscedasticity in real data with continuous explanatory variables than in unreplicated $2^k$ factorials.  In separate research, Gelfand (2015) studies 25 such data sets exhibiting ``significant'' heteroscedasticity, as determined by a statistically significant positive slope in a plot of the absolute values of residuals vs. fitted values.  Estimated standard deviation ratios between low- and high-variance regions are found to range from 1.1 to 64.4 (reduced to 15.6 with a possible outlier removed), with a median of 3.2 and a third quartile of 5.9.  Furthermore, Henrey and Loughin (2017) perform joint location-dispersion analysis of three oft-studied experiments, and find that two have apparent active dispersion effects.  In both of these, the effect estimate corresponds to a standard deviation ration of roughly 5.  Thus, our chosen values represent a range of realistic dispersion effect sizes, although the upper two levels might be somewhat extreme.}  Where two dispersion effects are present, we set $\Delta_A =\Delta_B$.

Because the variance of location-effect estimates changes depending on the magnitude of dispersion effects, we set the sizes of the location effects using the concept of EPower proposed by McGrath and Lin (2003).  Specifically, the magnitude of a location effect is defined so that the probability that it would be declared active using a $\sigma$-known $Z$ test with $\alpha=0.05$ is fixed at one of three levels: ``small'' (0.2), ``medium'' (0.5) or ``large'' (0.9).  Details on the calculations are given in McGrath and Lin (2003) and Zhang (2010).

For each combination of ${\cal L}, {\cal D}, \Delta_A$, and location-effect sizes, 1825 data sets are generated from model (\ref{heteroscedastic}) and analyzed using each of the four testing methods.  For brevity, we refer to the Lenth (1989) method as ``LEN89,'' the Berk and Picard (1991) method as ``BP91,'' the Box and Meyer (1986a) method as ``BM86,'' and the Loughin and Noble (1997) method as ``LN97.''  For each method, we record the number of times each of the 15 location effects is declared active.  We refer to this as the {\em rejection rate} (RR) for each effect.  In addition we compute three commonly-used summaries of these RRs:
\begin{itemize}
  \item {\em individual error rate} (IER), the average proportion of effects declared active among all inactive effects;
  \item {\em experimentwise error rate} (EER), the proportion of data sets in which at least one inactive effect is declared active; and
  \item {\em average power} (AP), the average proportion of effects declared active among all active effects;
\end{itemize}
These summaries have been commonly applied in the literature. For example, all three were used by Loughin and Noble (1997), Hamada and Balakrishnan (1998), and Ye et al. (2001), among others.  We an additional definition of power when there are two active location effects:
\begin{itemize}
  \item {\em joint power} (JP), the proportion of data sets in which {\em both} active effects are declared active.
\end{itemize}

As previous authors have found (e.g., Hamada and Balakrishnan 1998), not all analysis methods maintain their error rates at nominal levels using the default setting recommended by their respective authors.  Our findings, based on an initial set of 1825 simulations, are that LN97 and BP91 do have IER within simulation error of 0.05 using the recommended settings.  For LEN89, using the recalibrated critical values given in Loughin (1998) also results in acceptable control of IER.  However, BM86 needs to be recalibrated, meaning that a new posterior probability threshold must be chosen, above which an effect is declared active.  Using the approach described in Zhang (2010), we determined that a threshold of 0.3187 achieves acceptable IER.

Because the methods all use different theoretical foundations, they do not all exhibit the same EER when IER is held fixed.  Table~\ref{cab} gives the IER and EER of the calibrated methods based on 1825 simulated data sets.  It also shows the number of times 0, 1, 2, ... effects are falsely declared active.
\begin{table}[tbp]
\begin{center}
\caption { Performance of the four methods after calibration to IER=0.05. The values below the columns $0,1,\ldots,7$ are the proportions of simulations that declare that many effects as active when there are no location or dispersion effects.}\label{cab}\bigskip
\renewcommand{\arraystretch}{1.5}
\begin{tabular}{r|r|r|r|r|r|r|r|r|r|r}
\hline
\multirow{2}{*}{Method} & \multicolumn{8}{|c|}{Number of Declared Active Effects} & \multirow{2}{*}{IER} & \multirow{2}{*}{EER}\\\cline{2-9}
&0&1&2&3&4&5&6&$\geq7$&&\\
\hline
LEN89 & 0.586 & 0.200 & 0.099 & 0.067 & 0.028 & 0.009 & 0.008 & 0.003 & 0.051 & 0.414\\
\hline
BM86 & 0.526 & 0.292 & 0.098 & 0.042 & 0.019 & 0.010 & 0.009 & 0.004 & 0.051 & 0.474\\
\hline
BP91 & 0.541 & 0.264 & 0.116 & 0.055 & 0.018 & 0.005 & 0.002 & 0.000 & 0.050 & 0.459\\
\hline
LN97 & 0.726 & 0.140 & 0.037 & 0.032 & 0.012 & 0.008 & 0.014 & 0.039 & 0.049 & 0.274\\
\hline\hline
\end{tabular}
\end{center}
\end{table}

\section{Results}\label{SimResults}
In this section, we compare the four methods' estimated error rates and power.  Although we study cases with location effects of various magnitudes, results in terms of both error rates and power change in very predictable ways across different magnitudes of location effects.  We therefore focus attention on the case of location effects with medium EPower (theoretical power = 0.5), and observe changes in rejection rates across different magnitudes of dispersion effects.  Similarly, when there are two dispersion effects, cases where they differ in magnitude offer no special insights into the behavior of the analysis methods, so we focus on cases where the dispersion effects have the same magnitude.  More complete simulation results are available in Zhang (2010).

Where plots of rejection rates against dispersion-effect magnitude are presented, the values on the x-axis are deliberately left unscaled so that we may focus on more typical small-to-moderate sizes of dispersion effects.

\subsection{Error Rates}\label{ErrorRates}
First, Table~\ref{cab} shows that LN97 maintains a considerably different profile of effect-detection frequency than the other three methods even when there are no dispersion effects.  It tends to have a higher chance of detecting a large number of effects, which leads to a lower EER than other methods.  It is important to keep this context in mind when considering performance in the presence of active effects.

\begin{figure}
\begin{subfigure}{.5\textwidth}
  \centering
  \includegraphics[width=.9\linewidth]{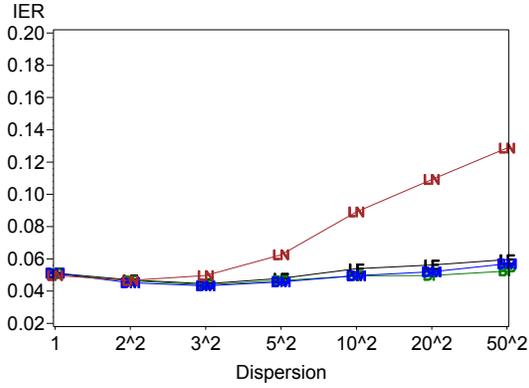}
  \caption{Individual error rates, ${\cal D}=\{A\}$}
  \label{IERL0D1}
\end{subfigure}%
\begin{subfigure}{.5\textwidth}
  \centering
  \includegraphics[width=.9\linewidth]{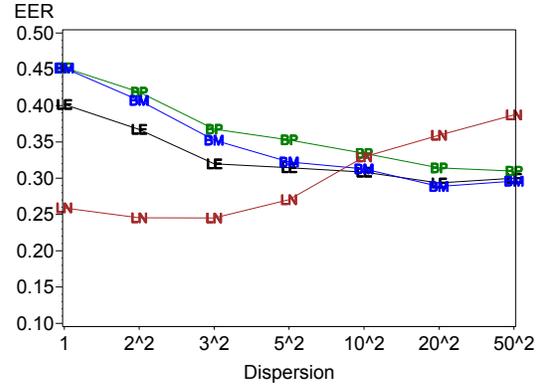}
  \caption{Experimentwise error rates, ${\cal D}=\{A\}$}
  \label{EERL0D1}
\end{subfigure}

\begin{subfigure}{.5\textwidth}
  \centering
  \includegraphics[width=.9\linewidth]{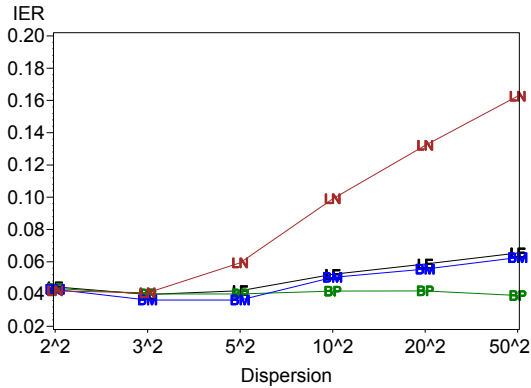}
  \caption{Individual error rates, ${\cal D}=\{A,B\}$}
  \label{IERL0D2}
\end{subfigure}%
\begin{subfigure}{.5\textwidth}
  \centering
  \includegraphics[width=.9\linewidth]{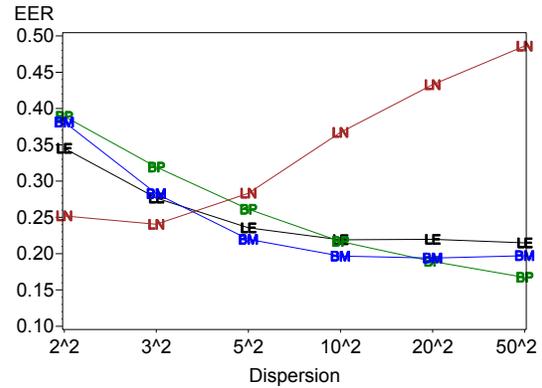}
  \caption{Experimentwise error rates, ${\cal D}=\{A,B\}$}
  \label{EERL0D2}
\end{subfigure}
\caption{Error rates of location-effect identification methods when there are dispersion effects of varying sizes.  Individual error rates are initially calibrated to 0.05.  `BM'=Box-Meyer, `BP'=Berk-Picard, `LE'=Lenth, `LN'=Loughin-Noble.}
\label{ErrorRate}
\end{figure}

Example IER plots against dispersion-effect magnitude are given in Figures \ref{IERL0D1} and \ref{IERL0D2}.  All four methods hold similar error rates near the nominal level when there are no dispersion effects or only very small ones.  All methods except LN97 maintain these error rates regardless of the presence of dispersion effects.  However, as the dispersion-effect magnitude grows, the IER for LN97 rises considerably.  The increase is relatively small as long as each $\Delta\le5$, but for more extreme dispersion effects the IER increases to well above 0.1.  Analogous results hold when there are active location effects, although in general the IERs decrease slowly as the number of active location effects increases.

The results for EER shown in Figures \ref{EERL0D1} and \ref{EERL0D2} reflect this same trend and show an interesting additional one.  The EER for LN97 increases sharply when the dispersion effects become larger, and the increase is greater when there are two dispersion effects than when there is one.  On the other hand, the EERs for the other three methods decrease, despite the relative stability of the IERs.  As discussed in Section \ref{DispEffs}, the increasing sizes of dispersion effects create increasing correlations among pairs of location-effect estimates.  The analysis methods apparently identify these pairs as jointly active or inactive with increasing relative frequency while maintaining approximately constant overall IERs.  Thus, a relatively smaller proportion of data sets experience at least one false rejection.

Results for cases with one or two active location effects show similar trends, although somewhat muted compared to Figures \ref{EERL0D1} and \ref{EERL0D2}.  For example, when there are two active location effects, the EER for LN97 does not surpass those for the other methods until $\Delta$ is at least $10^2$ in two cases and $20^2$ in one.

\subsection{Power}

\begin{figure}
\begin{subfigure}{.5\textwidth}
  \centering
  \includegraphics[width=.9\linewidth]{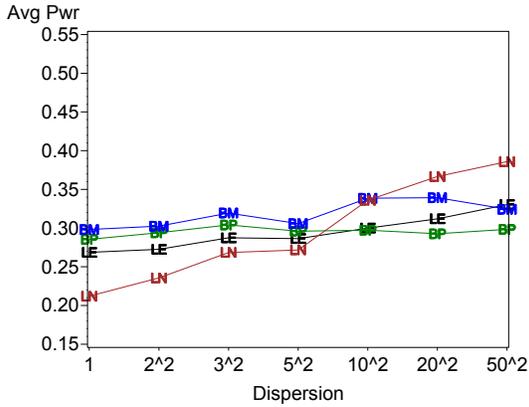}
  \caption{Average Power, ${\cal D}=\{A\},$ ${\cal L}=\{A\}$}
  \label{POWL1D1}
\end{subfigure}%
\begin{subfigure}{.5\textwidth}
  \centering
  \includegraphics[width=.9\linewidth]{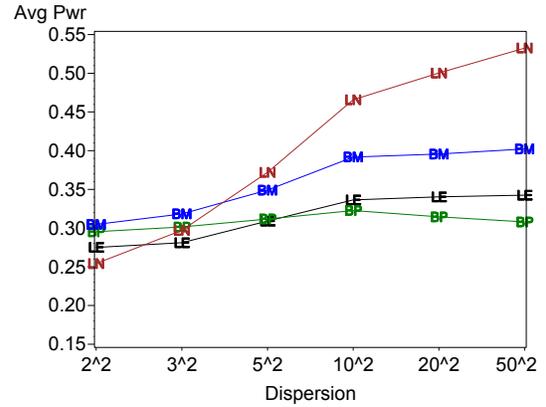}
  \caption{Average Power, ${\cal D}=\{A,B\},$ ${\cal L}=\{A\}$}
  \label{POWL1D2}
\end{subfigure}

\begin{subfigure}{.5\textwidth}
  \centering
  \includegraphics[width=.9\linewidth]{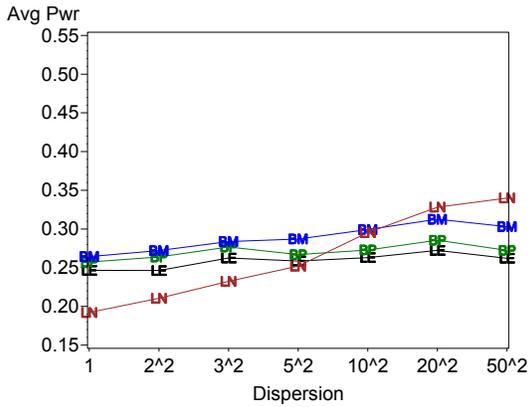}
  \caption{Average Power, ${\cal D}=\{A\},$ ${\cal L}=\{A,B\}$}
  \label{POWL2D1AB}
\end{subfigure}%
\begin{subfigure}{.5\textwidth}
  \centering
  \includegraphics[width=.9\linewidth]{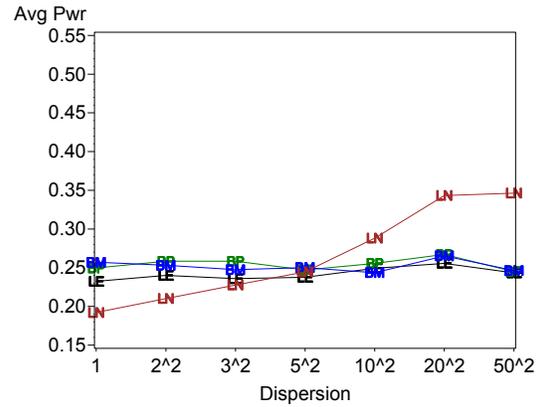}
  \caption{Average Power, ${\cal D}=\{A\},$ , ${\cal L}=\{B,AB\}$}
  \label{POWL2D1BAB}
\end{subfigure}
\caption{Average power of location-effect identification methods when there are dispersion effects of varying sizes.  Theoretical EPowers are initially calibrated to 0.50.  `BM'=Box-Meyer, `BP'=Berk-Picard, `LE'=Lenth, `LN'=Loughin-Noble. Note that a 95\% confidence interval around an estimate of 0.30 is approximately $\pm 0.02$.}
\label{AveragePower}
\end{figure}

Average power for a sampling of cases is shown in Figure \ref{AveragePower}.  The most notable feature on all of these cases is that the average power of LN97 increases as the disparity in dispersions increases.  This is contrasted against little to no increase in average power for other three analysis methods.  Among these three, BM86 sometimes enjoys a slight advantage in power, particularly when there are two dispersion effects.

Comparing across cases, LN97 shows a much starker increase in power with dispersion-effect magnitude when there are two dispersion effects than when there is one.  This pattern is evident regardless of which factor contains the active location effect, and is consistent with this method's elevated error rates with two dispersion effects.  The other three methods also experience power increases in these cases, albeit of a much smaller magnitude.  In no case does the estimated power reach the EPower level to which the location effect sizes were calibrated.  This is not surprising, because the EPowers are based on the known error variance and a normal sampling distribution.  In unreplicated experiments, the ability to estimate process variance accurately is hampered by the lack of data and the uncertainty associated with trying to distinguish between active and inert effects.

\begin{figure}
\begin{subfigure}{.5\textwidth}
  \centering
  \includegraphics[width=.9\linewidth]{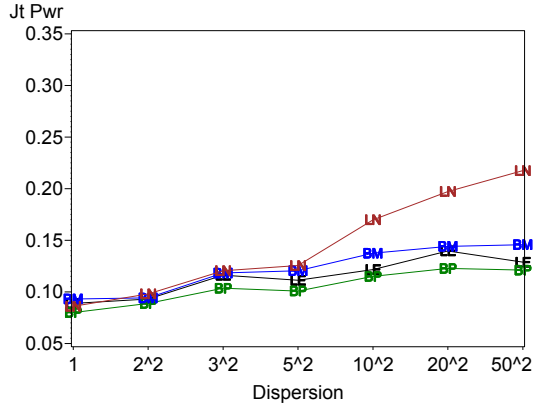}
  \caption{Joint Power, ${\cal D}=\{A\},$ ${\cal L}=\{A,B\}$}
  \label{JPL2D1AB}
\end{subfigure}%
\begin{subfigure}{.5\textwidth}
  \centering
  \includegraphics[width=.9\linewidth]{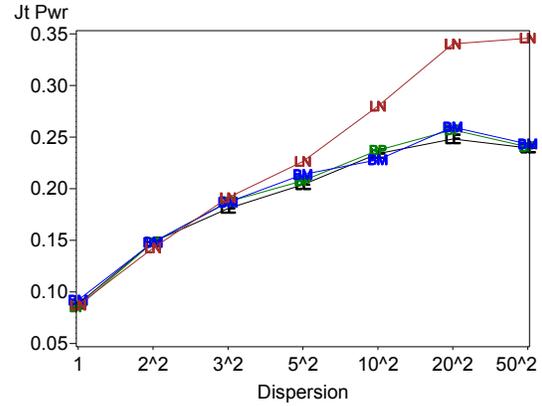}
  \caption{Joint Power, ${\cal D}=\{A\},$ ${\cal L}=\{B,AB\}$}
  \label{JPL2D1BAB}
\end{subfigure}
\caption{Joint power of location-effect identification methods when there are two location effects and one dispersion effect.  Theoretical powers are initially calibrated to 0.50 per location effect.  `BM'=Box-Meyer, `BP'=Berk-Picard, `LE'=Lenth, `LN'=Loughin-Noble. }
\label{JtPow}
\end{figure}

Plots of the joint power to detect both location effects are given in Figure \ref{JtPow} for cases where there is one dispersion effect and two location effects.  The joint power is clearly higher when the dispersion effect occurs at the intersection of the two location effects (Figure \ref{JPL2D1BAB}) than when it does not (Figure \ref{JPL2D1BAB}), and it increases considerably as the size of the dispersion effect increases.  This is a clear result of the positive correlation induced on the location-effect estimates by a spurious dispersion effect at their intersection.  While this may seem at first to be a welcome result---a way to ``boost power for free''---note that this apparent boost in power only occurs when the sign of $\log(\Delta_A)$ matches the sign of $\beta_B \beta_{AB}$.  If they had been of opposite signs, the power would have decreased, rather than increased, with increasing magnitude of $\Delta_A$.

\begin{figure}[t]
 \begin{subfigure}{.5\textwidth}
  \centering
  \includegraphics[width=.9\linewidth]{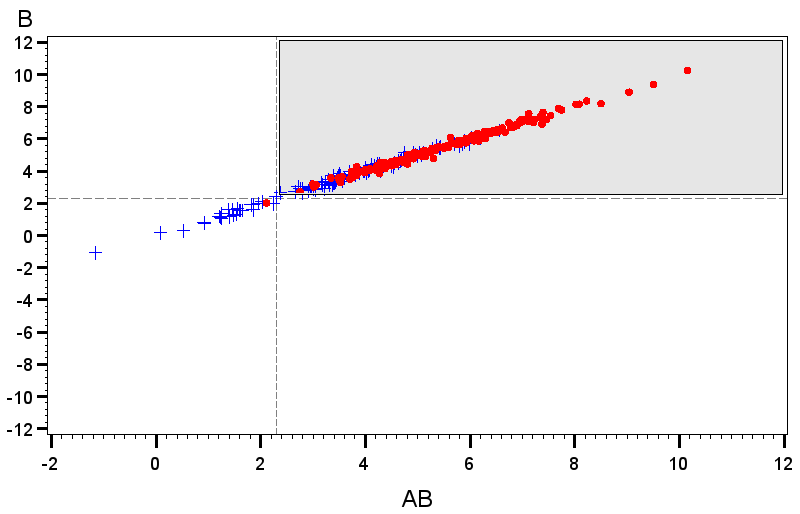}
  \caption{$\beta_B \beta_{AB} > 0$}
  \label{BAB+}
\end{subfigure}
 \begin{subfigure}{.5\textwidth}
  \centering
  \includegraphics[width=.9\linewidth]{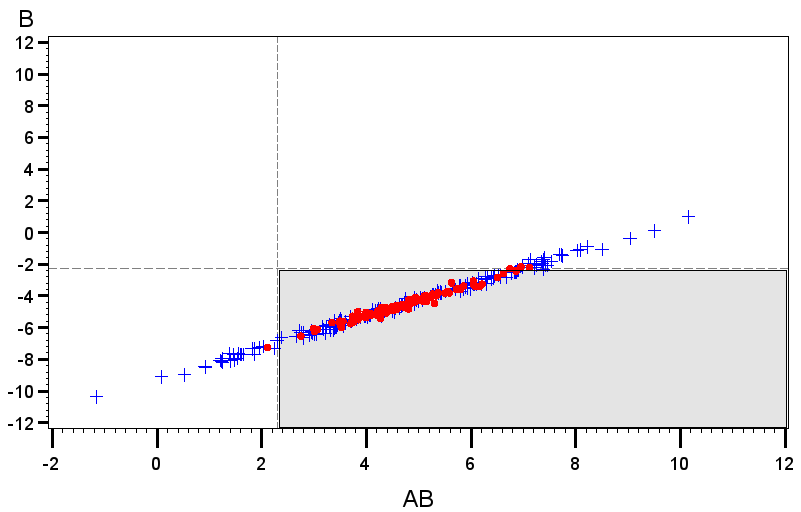}
  \caption{$\beta_B \beta_{AB} < 0$}
  \label{BAB-}
\end{subfigure}
\caption{$\hat{\beta}_B$ vs $\hat{\beta}_{AB}$ in the case ${\cal D}=\{A\}, {\cal L}=\{B,AB\}$ with $\Delta_A=400$.  The shaded region is the theoretical rejection region for declaring both location effects jointly active. A red `$\bullet$' represents that both $B$ and $AB$ are declared active; a blue `$+$' indicates otherwise.}\label{BAB}
\end{figure}

\begin{figure}[t]
 \begin{subfigure}{.5\textwidth}
  \centering
  \includegraphics[width=.9\linewidth]{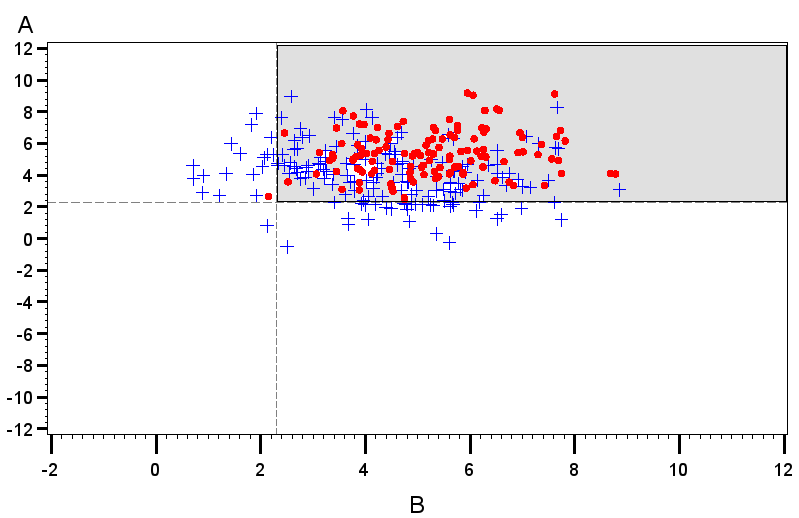}
  \caption{$\beta_B \beta_AB > 0$}
  \label{AB+}
\end{subfigure}
 \begin{subfigure}{.5\textwidth}
  \centering
  \includegraphics[width=.9\linewidth]{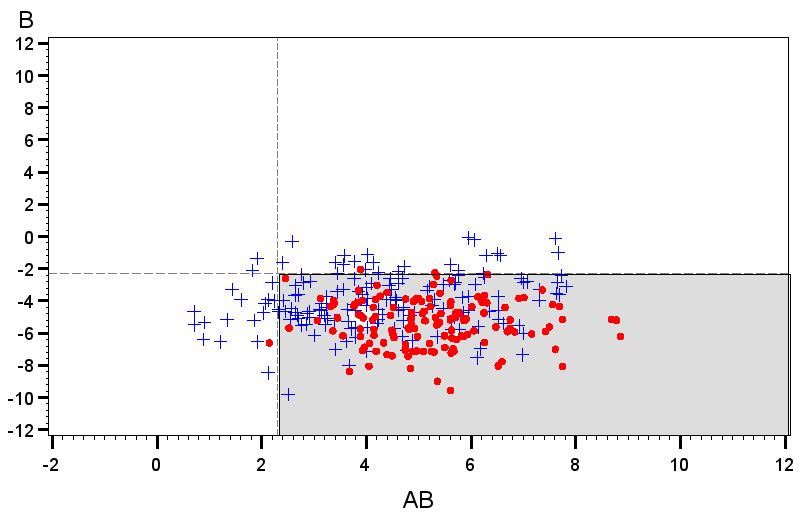}
  \caption{$\beta_B \beta_AB < 0$}
  \label{AB-}
\end{subfigure}
\caption{$\hat{\beta}_A$ vs $\hat{\beta}_{B}$ in the case ${\cal D}=\{A\}, {\cal L}=\{A,B\}$ with $\Delta_A=400$.  The shaded region is the theoretical rejection region for declaring both location effects jointly active. A red `$\bullet$' represents that both $A$ and $B$ are declared active; a blue `$+$' indicates otherwise.}\label{AB}
\end{figure}

To further explore this property, Figure~\ref{BAB} shows observed distributions of $\hat{\beta}_B$ and $\hat{\beta}_{AB}$ in the somewhat extreme case where $\Delta_A=400$ and the EPowers for the two location effects effects are both 0.9.  In the right plot, both location effects are positive, while on the left, $\beta_B<0$.  Figure~\ref{AB} shows the analogous plot for the case where  ${\cal D}=\{A\}, {\cal L}=\{A,B\}$.  The the theoretical rejection regions on which the EPowers are based are given by the shaded regions in each plot.  Different plotting symbols indicate whether the pair of effects was jointly declared active.  Note that the actual rejection region varies across data sets in the simulations, so not all points within the theoretical rejection region are identified as active and not all points outside it are declared inactive.

Figure~\ref{BAB} shows the strong correlation between location effect estimates that is induced by the extreme dispersion effect.  It also shows that the fraction of points that falls within the rejection region is substantially higher when the correlation causes the distribution of points to align with the position of this region than when the distribution cuts across the corner of it.  Thus, joint power is much higher when the sign of $\log(\Delta_A)$ matches the sign of $\beta_B \beta_{AB}$.  On the other hand, in Figure~\ref{AB} the two location effect estimates are uncorrelated, because the interaction of these terms does not match the factor for the dispersion effect.  Thus, the fraction of the distribution that falls witin the rejection region is the same, regardless of the signs of the true location effects.

\begin{figure}[t]
    \includegraphics[width=.9\linewidth]{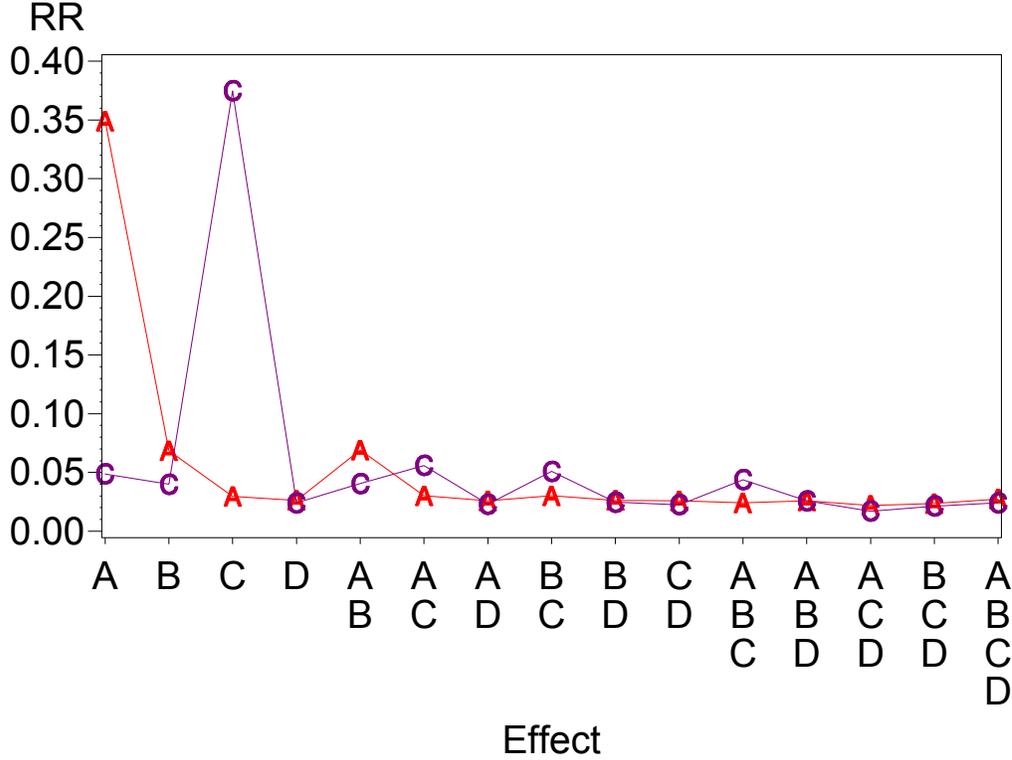}
    \caption{Rejection rates for each effect when BM86 is used on model ${\cal D}=\{A,B\}, \Delta_A=\Delta_B=5^2$ with either ${\cal L}=\{A\}$ (red) or ${\cal L}=\{C\}$ (purple) with medium EPower.}\label{RRbm}
    \end{figure}

The influence of dispersion effects on tests for each different location effect is shown in Figure \ref{RRbm}, using BM86 as a representative of the methods.  This plot shows the rejection rate for each individual effect when BM86 is applied to the model ${\cal D}=\{A,B\}, \Delta_A=\Delta_B=5^2$ with either ${\cal L}=\{A\}$ or ${\cal L}=\{C\}$ set to medium EPower.  We see in each case that the active location effect has a much higher rejection rate than the inactive effects, as expected.  However, we also see that the error rates for inert location effects associated with (1) the dispersion effects, (2) the interaction of dispersion effects ($AB$), and (3) the interaction of location and dispersion effects ($AC,BC,ABC$ in the second case) are all slightly elevated relative to the baseline set by the remaining effects.  Indeed, the location effect at $B$ for ${\cal L}=\{A\}$ experiences the dual increase due to being associated both with a dispersion effect and with a location-dispersion interaction through $A$ and $AB$.  The location effect at $AB$ for ${\cal L}=\{A\}$ experiences a similar increase.  In all cases where the ``A'' and ``C'' lines appear different for a given location effect, this difference is statistically significant at the 0.01 level or lower.  Thus, it is clear that interacting with a dispersion effect causes a location effect to experience a higher rejection rate.

\section{Discussion and Conclusions}\label{concl}
This paper considerably expands our understanding of the impact of dispersion effects on location-effect estimation in unreplicated $2^k$ (fractional) factorial designs.  Pan (1999) conducted a simulation consisting of a single case with one dispersion effect set to $\Delta=7^2$, using only Len89 for the analysis of location effects.  He found that there was little impact on estimated RR for any of the effects, which we corroborate here.  However, our finding that the LN97 is somewhat affected by unidentified dispersion effects indicates that Pan's result does not hold equally for all methods.  Among the four methods that we considered, BM86 may enjoy a very slight advantage in terms of power when dispersion effects are added to the data.  However, the advantage is small, and the simplicity of Len87 is appealing.

The heteroscedasticity-induced correlation has some side effects that are documented here.  While the correlation has minimal effect on the marginal power to detect a real effect, the joint power of members of an interaction triple is radically changed.  Depending on whether the direction of the dispersion effect agrees or disagrees with the sign of the product of the location effects, joint power may be either increased or decreased by the dispersion effect.  In problems where there is also interest in identifying dispersion effects, this can create a feedback loop.  In the case where the sign on the product agrees with the dispersion effect, there is increased probability that both location effects are left out of the model together.  This artificially enhances the size of the dispersion effect, increasing the chance that it is detected but falsely inflating its estimated magnitude as well (McGrath and Lin 2001a).   On the other hand, when the sign on the product disagrees the situation is less concerning: the probability of failing to detect both active location effects is smaller than usual, so there is a reduced chance of an impact on the dispersion effect identification.  When it does occur, however, the apparent magnitude of the dispersion effect is reduced, making it harder to detect.

As with any simulation study, there are certain limitations to these results.  Since these four analysis methods were hand-picked from among many possible approaches---albeit with reason as discussed in Section \ref{LocEffs}---one should be careful not to infer anything about how other methods might or might not be impacted by dispersion effects.  This is particularly true considering the adverse effects on LN97, which suggests that different methods do not all respond in the same way.  Similarly, we are limited to drawing conclusions about the cases actually studied here---one or two dispersion effects and zero, one, or two location effects---and can make only educated guesses regarding trends for other cases.   Also, since we only studied the $2^4$ case explicitly, one should be tentative about extending the results to other experiment sizes.  However, simulations conducted by Balakrishnan and Hamada (1998) indicate that there is nothing different about the size of the problem that changes how well different methods perform, and there is nothing about the mathematics of the problem that depends on the size.

In conlusion, when an analyst wants to identify location effects and has no interest in dispersion effects, the results in this paper provide comfort that several standard methods of location-effect identification are relatively robust in the presence of dispersion effects.  However, when there may also be interest in understanding the dispersion effects in the data, there are better methods for identifying both effect types than the sequential application of separate location- and dispersion-effect estimation methods.  In particular, Henrey and Loughin (2017) derive a new information criterion for heteroscedastic normal linear models and use it in a model-averaging context to simultaneously identify {\it both} location and dispersion effects.  They demonstrate that this procedure can identify both location and dispersion effects with greater accuracy than a combined sequential procedure consisting of Len87 for location effects and the modified Harvey (1976) method suggested by Brenneman and Nair for dispersion effects.



\section{Acknowledgments}
This work was supported by the National Science and Engineering Research Council of Canada.

\pagebreak

\section*{References}


\vskip.2in
\noindent\parbox[t]{6in}{\setlength{\baselineskip}{14pt} Benski, H.C. (1989).  ``Use of a Normality Test to Identify Significant Effects in Factorial Designs''. \it Journal of Quality Technology\rm, 21, pp.~174--178.}

\vskip.2in
\noindent\parbox[t]{6in}{\setlength{\baselineskip}{14pt} Bergman, B. and Hyn\'{e}n, A. (1997).  ``Dispersion Effects from Unreplicated Designs in the \mbox{$2^{k-p}$} Series''. \it Technometrics\rm, 39, pp.~191--198.}

\vskip.2in
\noindent\parbox[t]{6in}{\setlength{\baselineskip}{14pt} Berk, K.N. and Picard, R.R. (1991).  ``Significance Tests for Saturated Orthogonal Arrays''. \it Journal of Quality Technology\rm, 23, pp.~79--89.}

\vskip.2in
\noindent\parbox[t]{6in}{\setlength{\baselineskip}{14pt} Bissell, A.F. (1992).  ``Mean squares in Saturated Factorial Designs Revisited''. \it Journal of Applied Statistics\rm, 19, pp.~351--366.}


\vskip.2in
\noindent\parbox[t]{6in}{\setlength{\baselineskip}{14pt} Box, G.E.P., Hunter, W.G., and Hunter, J.S. (2005).
\it Statistics for Experimenters, 2nd ed.\rm, John Wiley \& Sons, Hoboken, NJ.}

\vskip.2in
\noindent\parbox[t]{6in}{\setlength{\baselineskip}{14pt} Box,
G.E.P. and Meyer, R.D. (1986a).  ``An analysis for unreplicated fractional factorials''. \it Technometrics\rm, 28, pp.~11–18.}

\vskip.2in
\noindent\parbox[t]{6in}{\setlength{\baselineskip}{14pt} Box,
G.E.P. and Meyer, R.D. (1986b). ``Dispersion Effects from Fractional Designs''. \it Technometrics\rm, 28, pp.~19--27.}

\vskip.2in
\noindent\parbox[t]{6in}{\setlength{\baselineskip}{14pt} Brenneman,
W.A. and Nair, V.N. (2001). ``Methods for Identifying Dispersion
Effects in Unreplicated Factorial Experiments: A Critical Analysis and
Proposed Strategies''. \it Technometrics\rm, 43, pp.~388--405.}






\vskip.2in
\noindent\parbox[t]{6in}{\setlength{\baselineskip}{14pt} Cook,
R.D. and Weisberg, S. (1983). ``Diagnostics for Heteroscedasticity in
Regression''. \it Biometrika\rm, 70, pp.~1--10.}

\vskip.2in
\noindent\parbox[t]{6in}{\setlength{\baselineskip}{14pt} Daniel, C. (1959). ``Use of Half-Normal Plots in Interpreting Factorial Two-Level Experiments''. \it Technometrics\rm, 1, pp.~311--341.}

\vskip.2in
\noindent\parbox[t]{6in}{\setlength{\baselineskip}{14pt} Davidian, M and Carroll, R.J. (1987). ``Variance Function Estimation''. \it Journal of the American Statistical Association\rm, 82, pp.~1079--1091.}


\vskip.2in
\noindent\parbox[t]{6in}{\setlength{\baselineskip}{14pt} Dean, A. and Lewis, S. (Eds.) (2006). {\it Screening: Methods for Experimentation in Industry, Drug Discovery, and Genetics\rm}, Springer, London.}

\vskip.2in
\noindent\parbox[t]{6in}{\setlength{\baselineskip}{14pt} Dong, F. (1993). ``On the Identification of Active Contrasts in Unreplicated Fractional Factorials''. \it Statistica Sinica\rm, 3, pp.~209--218.}




\vskip.2in
\noindent\parbox[t]{6in}{\setlength{\baselineskip}{14pt} Gelfand, S. (2015).  ``Understanding the impact of heteroscedasticity on the predictive ability of modern regression methods''.  MSc. Thesis, Simon Fraser University, Burnaby, BC, Canada.}

\vskip.2in
\noindent\parbox[t]{6in}{\setlength{\baselineskip}{14pt} Grego, J.M., Lewis, J.F., and Craney, T.A. (2000).  ``Quantile Plots for Mean Effects in the Presence of Variance Effects for $2^k$ Designs''.  {\it Communications in Statistics---Computation and Simulation}, 29, pp.~1109--1133.}


\vskip.2in
\noindent\parbox[t]{6in}{\setlength{\baselineskip}{14pt} Hamada,
M. and Balakrishnan, N. (1998). ``Analyzing Unreplicated Factorial
Experiments: A Review with Some New Proposals''. \it Statistica Sinica\rm, 8, pp.~1--41.}

\vskip.2in
\noindent\parbox[t]{6in}{\setlength{\baselineskip}{14pt} Harvey, A.C. (1976). ``Regression Models with Multiplicative Heteroscedasticity'' {\it Econometrica}, 44, pp.~461--465.}	

\vskip.2in
\noindent\parbox[t]{6in}{\setlength{\baselineskip}{14pt} Henrey, A.J. and Loughin, T.M. (2017). ``Joint Identification of Location and Dispersion Effects in Unreplicated Two-Level Factorials'' {\it Technometrics}, 59, pp.~23--35.}	



\vskip.2in
\noindent\parbox[t]{6in}{\setlength{\baselineskip}{14pt} Juan, J. and Pe$\tilde{\hbox{n}}$a, D. (1992). ``A Simple Method to Identify Significant Effects in Unreplicated Two-level Factorial Designs''. {\it Communications in Statistics - Theory and Methods}, 21, pp.~1383--1403.}


\vskip.2in
\noindent\parbox[t]{6in}{\setlength{\baselineskip}{14pt} Lenth,
R.V. (1989). ``Quick and Easy Analysis of Unreplicated Factorials''. \it
Technometrics\rm, 31, pp.~469--473.}

\vskip.2in
\noindent\parbox[t]{6in}{\setlength{\baselineskip}{14pt} Loughin, T.M. (1998). ``Calibration of the Lenth Test for Unreplicated Factorial Designs''. {\it Journal of Quality Technology}, 30, pp.~171--175.}


\vskip.2in
\noindent\parbox[t]{6in}{\setlength{\baselineskip}{14pt} Loughin, T.M. and Noble, W. (1997). ``A Permutation Test for
Effects  in an Unreplicated Factorial Design''. {\it Technometrics}, {39},  pp.~180--190.}


\vskip.2in
\noindent\parbox[t]{6in}{\setlength{\baselineskip}{14pt} McGrath,
R.N. and Lin, D.K. (2001a). ``Confounding of Location and Dispersion
Effects in Unreplicated Fractional Factorials''. \it Journal of Quality
Technology\rm, 33, pp.~129--139.}

\vskip.2in
\noindent\parbox[t]{6in}{\setlength{\baselineskip}{14pt} McGrath,
R.N. and Lin, D.K. (2001b). ``Testing Multiple Dispersion Effects in
Unreplicated Fractional Factorial Designs''. \it Technometrics\rm,
43, pp.~06--414.}

\vskip.2in
\noindent\parbox[t]{6in}{\setlength{\baselineskip}{14pt} McGrath, R.N. (2003).	
``Separating Location and Dispersion Effects in Unreplicated Fractional Factorial Designs''.
{\it Journal of Quality Technology}, 35, pp.~306--316.}	







\vskip.2in
\noindent\parbox[t]{6in}{\setlength{\baselineskip}{14pt} Pan,
G. (1999).  ``The Impact of Unidentified Location Effects on Dispersion
Effect Identification from Unreplicated Factorial Designs''. \it
Technometrics\rm, 41, pp.~313--326.}

\vskip.2in
\noindent\parbox[t]{6in}{\setlength{\baselineskip}{14pt} Pan, G and Taam, W (2002).
``On Generalized Linear Model Method for Detecting Dispersion Effects in Unreplicated Factorial Designs''.
{\it Journal of Statistical Computation and Simulation}, 72, pp.~431--450.}


\vskip.2in
\noindent\parbox[t]{6in}{\setlength{\baselineskip}{14pt} Rao,
C.R. (1970). ``Estimation of Heteroscedastic Variances in Linear
Models''. \it Journal of the American Statistical Association\rm,
65, pp.~161--172.}




\vskip.2in
\noindent\parbox[t]{6in}{\setlength{\baselineskip}{14pt} Wu, C.F.J. and Hamada, M.S. (2000). {\it Experiments: Planning, Analysis, and Parameter Design Optimization}. John Wiley and Sons, Hoboken, NJ.}

\vskip.2in
\noindent\parbox[t]{6in}{\setlength{\baselineskip}{14pt} Ye, K., and Hamada, M. (2000).  ``Critical Values of the Lenth Method for Unreplicated Factorial Designs''. {\it Journal of Quality Technology}, 32, pp.~57--66.}

\vskip.2in
\noindent\parbox[t]{6in}{\setlength{\baselineskip}{14pt} Ye, K.Q., Hamada, M., and Wu, C.F.J. (2001).  ``A Step-Down Lenth Method for Analyzing Unreplicated Factorial Designs''. {\it Journal of Quality Technology}, 33, pp.~140--152.}

\vskip.2in
\noindent\parbox[t]{6in}{\setlength{\baselineskip}{14pt} Zahn, D.A. (1975).  ``Modifications and Revised Critical Values for the Half-Normal Plot''. {\it Technometrics}, 17,  pp.~189--200.}

\vskip.2in
\noindent\parbox[t]{6in}{\setlength{\baselineskip}{14pt} Zhang, Y. (2010).  ``A Comparison of Location Effect Identification Methods for Unreplicated Fractional Factorials in the Presence of Dispersion Effects''.  MSc. Thesis, Simon Fraser University, Burnaby, BC, Canada.}

\end{document}